# Statistical properties of geomagnetic measurements as possible precursors for magnetic storms

Andrés R. R. Papa[1,2,*] and Lilian P. Sosman[2]

[1]Observatório Nacional, Rua General José Cristino 77, São Cristóvão,
Rio de Janeiro, 20921-400 RJ, BRASIL
[2]Instituto de Física, Universidade do Estado do Rio de Janeiro, Rua São Francisco Xavier 524, Maracanã,
Rio de Janeiro, 20550-900 RJ, BRASIL

**Summary**

Records of geomagnetic measurements have been analyzed looking for evidences of possible precursors for magnetic storms. With this objective, the main magnetic storms in the period 1998-2002 have been located in Dst index record. Periods immediately before storms and periods well before them were studied by applying a method recently introduced in the literature. Statistical properties of both types of periods have been compared. One of the compared quantities was the slope of the power laws that have been found for some relevant distributions. A systematic deviation between slope distributions was found. This might be the fingerprint of a non self-organized component in records. There has also been found a correlation between slope values and the corresponding storm intensities, which could serve as a probabilistic approach to magnetic storms forecasting. Data is from the low latitude Vassouras Magnetic Observatory.

---

* E-mail addresses: papa@on.br and sosman@uerj.br

**Introduction**

Geomagnetic storms and geomagnetic reversals (Ponte-Neto and Papa, 2006) are the most striking phenomena in the magnetic field that we can measure at the Earth's surface. Geomagnetic storms are periods from one to three days during which the magnetic field born at the Sun surface and atmosphere of the Earth presents strong variations that, among other things, seriously affect telecommunication transmissions. Some evidences suggest that during magnetic storms the number of heart attacks increases in relation to calm periods (Halberg et al., 2001). Very recently (November 2003), around 50000 Swedish people suffered an energy blackout during a short period of time caused by a strong magnetic storm. This was also the case during the year of 1989 (but lasting more than nine hours) for the region of Quebec, in Canada. For some scientific and technical applications (for example, measurements of the crustal magnetic field for prospecting purposes, and directional drilling) the Earth's magnetic field is supposed to be known and used as a reference (Gleisner et al., 2005). Any variation or deviation from undisturbed values may be the source of unwanted errors. There are other methods of prospecting (MT, for example) in which the occurrence of magnetic perturbations (far from measurement points to accomplish the condition of plane wave front) is desired and useful.

All the reasons mentioned above justify the search for methods that allow predict magnetic storms at medium and long terms. But as in the case of other catastrophic events (earthquakes, for example) that affect human activities and threat human life there do not exist yet methods to accomplish this task in a reliable manner.

Magnetic storms have been advanced as possible candidates to precursors of other geophysical phenomena like earthquakes (Enescu, 2005; Kushwah, 2004). However, this is a very recent proposal and it is a point of current discussion.

Many efforts have been devoted to study statistical properties of magnetic measurements at the Earth's surface and satellites during the last five decades. We can mention, for example, a pioneering power spectrum study for the magnetic field around the world (Alldredge et al., 1963), and more recently, an intermittency analysis of geomagnetic series (Bolzan et al, 2005), a detrended fluctuation analysis on Sym-H index record (Wanliss, 2005), and a statistical study of direct magnetic field measurements (Papa et al., 2006).

The search for geomagnetic storms precursors can be done from some Earth grounded systems as, for example, muon detectors (Jansen et al, 2001) and geomagnetic observatories. There are also some Earth and satellites systems to study visible light and X-ray burst during solar flares (Yermolaev and Yermolaev, 2003). However, as well as in the case of the possibility of magnetic storms being earthquakes precursors, this possibility requires further scrutiny (Yermolaev and Yermolaev, 2006). For a comprehensible review, on scientific models for space weather developments, see the work by Lathuillère et al. (2002).

The inspiration for the present work came mostly from the works by Wanliss (2005) and Papa et al. (2006). However our work follows well defined different lines. Wanliss developed an extensive study (from 1981 to 2000) of fractal properties of the Sym-H index dividing the record in quiet and active intervals (based on Kp values), while Papa et al. concentrated their efforts in studying a short period of time (October 2000) through a Fourier analysis of magnetic measurements that diminish some of the problems that are normally faced because the non-stationary character of those series. Here we compare fractal properties of relatively calm periods, some of them near magnetic storms' beginnings and some others far (see below) from magnetic storms. We also compare those properties with the corresponding magnetic storms intensities. A possibility for medium term (~ three days) magnetic storm probabilistic forecasting emerges.

**Data analysis and Preparation**

The Dst and Sym-H indexes are essentially the same thing except for the time resolution (one hour for Dst and one minute for Sym-H). However, it is Dst that is widely used as a magnetic storm revealer. It is often taken the value –50 nT of that index as the threshold to consider that a magnetic storm is taken place (Maltsev, 2003). Figure 1 presents the variation of the Dst index for the month of October 2000. The data was obtained at the World Data Centre for Geomagnetism, Kyoto (WDC). The unique general feature for storm periods that we can note from Figure 1 and other monthly periods (not shown) is that does not exist any general feature for those periods. There are, however, two characteristics that are present in a great number of magnetic perturbation cases: a sudden decrease in Dst values (during a few hours) and a slow recovering to values near zero (i.e., "normal" values). We based our data selection and preparation on those features.

To avoid problems resulting from the highly non-stationary character of Dst series there are two policies that are normally used. The first obtains robust averages through studies of long periods of time (Wanliss, 2005). The second, assuming a quasi-stationary character for the series, studies short periods of time. In this work our choice was the last one.

The preparation of data consists in two set of relatively simple steps that we describe now. The value –50 nT of the Dst index was used as the threshold criteria to locate the beginning of magnetic storms. We have located the date and hour of the beginning of all the magnetic storms in the period 1998–2002, obtaining around 130 occurrences (~ two by month). The last value greater or equal to –10 nT before the –50 nT value that marks the storm beginning was considered to be the beginning of the transient before storms. Not all the periods with values greater or equal to –50 nT classify as storm with those criteria because we have not considered transients greater than 10 hours. The corresponding "storm" does not enter our record. They are walkings of Dst in far from zero values or rests of previous storms. The mean value for the jump period from a value greater than –10 nT values below –50 nT corresponds to around 6 hours, and this value was used for all the storms. We have also eliminated from our record the periods of Dst less than –50 nT too close to previous storms because would not have the complete period to be studied (see below). The final result was a set with around 50% of the periods with some point at Dst values lesser than –50 nT, giving 60 magnetic storms. Table 1 presents storms that entered the set to be studied.

Once determined from Dst records the periods of interest we have applied a recent methodology (Papa et al., 2006) to two types of periods before storms. We have investigated two time windows of three days: the three consecutive days before the transient beginning and the three consecutive days before them. In a few words the methodology consists in applying a Fourier transform to the selected data, filtering it through a second order Butterworth filter, returning the transform to the time domain and, finally, calculate the difference ΔH between original and filtered data (see the work by Papa et al. (2006) for details). The ΔH distribution is the object of study. When we apply this methodology to relatively long periods (one month) we obtain distributions for ΔH that obey to power laws. Power laws have the form:

$$f(x) = k \cdot x^d \qquad (1)$$

where *f(x)* is the frequency distribution on the variable *x*, *k* is a proportionality constant and *d* is the exponent of the power law (power laws have the property of appearing as straight lines in log-log plots).

The distribution of ΔH for the month of October 2000 is shown in Figure 2. There is a two-regime power law with a knee between both regimes always at values between 10 nT and 20 nT. Applying the methodology to three days periods is a dangerous strategy because we dramatically diminish (by a factor of 10) the number of data (giving a worse statistics, therefore). A typical result for a three days period is shown in Figure 3a. Now, the second power law regime is completely lost. We have performed our analysis on the slope of the power law regime here present. As a consequence of the statistical worsening we have established a threshold for the accuracy of slopes. Slopes with relative error greater than 20% were dropped out of the set under consideration. We have included in the study, at the end, around 45 storms.

The amplitude-frequency distribution of the direct Fourier transform follows also a power law distribution with slopes values around one. A typical result can be seen in Figure 2b. We have studied in parallel their distribution.

Before presenting the Results and Discussion section there are a few comments that we would like to do on data analysis and preparation. Because of the highly irregular behavior of Dst it was not possible to establish an automatic (man free) computer procedure to select the periods to be considered. In this sense, the results here presented are a hand-made product. They were individually analyzed case by case. A similar situation was faced when analyzing the slopes of $\Delta H$ distributions. This can be a serious difficulty to extend the study to longer periods and to do studies with validation purposes.

**Results and Discussion**

It is well known that distributions of magnetic field measurements (Papa et al, 2006) as well as values of some geomagnetic indexes (Wanliss, 2005) present power laws or, in other words, might be compatible with some fractal system.

In Figure 4 we present the time dependence of the slope in $\Delta H$ and transform amplitude distributions for the period 1998-2002. As can be seen in panel *a* of Figure 4, while there does not exist a one-to-one correlation between slopes for periods near and far from storms, there seems to be a common long term trend (increasing as time goes by) in both slope values. No trivial trend is observed in panel *b* of Figure 4. These results can be partially corroborated in Figure 5 where correlation graphs between slopes in both types of periods are shown. In panel *a* of

Figure 5 a concentration of points around the straight-line *y=x* is clear (independently of the used scales). In Figure 5b this concentration is not observed. Figure 6 shows the slope distributions for both types of three days intervals during the period 1998-2002. The distribution in Figure 6a shows a trend to higher slope values (lower in absolute value) for periods closer to storms. This tendency has also been noted in studies on the Kp index (Dias et al., 2006). The distributions in Figure 6b are statistically indistinguishable as shown by a set of tests that we have performed (considering them as single peak distributions). It is remarkable, however, the presence of double peak distributions for both kinds of periods (but this has to be confirmed by studies on larger sets of data). Two-peak distributions in some fractal properties of Sym-H index have also been found (Wanliss, 2005). The tendency presented in Figure 6a could be the fingerprint of a non self-similar component in geomagnetic records, which means that effective medium term forecasting might be a no so hard task.

Finally, in Figure 7, we present the dependence of slopes on the corresponding magnetic storm intensity. A well-behaved forbidden region is noted in Figure 7a while a not so well delimited forbidden region is observed in Figure 7b. For forecasting purposes forbidden regions are useful. While they can no tell what is going to happen, at least they can tell us what kind of events are not going to happen or the likelihood of a given event. Let us briefly exemplify those affirmations with base in Figure 7a. It can be deduced from Figure 7a that if the slope for the period near storm is below the value $-2.5$ then the corresponding storm will never have intensity greater than $-150$ nT. On the other hand, we can see from the same figure that, if the slope for the nearest period is between $-1.75$ and −1.25 there is around 30% of probability for the corresponding storm to be more intense than −150 nT. The 30% probability was obtained by noting that there are five points (circles) in the rectangle formed by the values d = −1.75 and −1.25 and storm-intensity = −150 nT and −325 nT, while there are ten points (also circles) in the rectangle formed by the same values of d and storm-intensity = −150 nT and −50 nT. It should also be noted from Figure 7a that, given that there are not slope values above −0.9, the periods near storms define a more restrictive forbidden region (circles) if compared with the forbidden region defined by periods far from storms (squares). This is intuitively clear if we believe that there is some effective change in slopes from one period type to the other as it seems to be the case: the closer the period to the storm the more accurate description it provides on the subsequent storm.

We consider the results presented in Figure 7 the main finding of the present work.

**Conclusions**

We have studied the change in fractal properties of magnetic data from our low latitude Vassouras Magnetic Observatory and their relation with storms intensities. There are several things to be learnt from our study. There are periods with Dst values > −50 nT that accomplish the requirements of having a sudden decrease in a few hours and a relaxation for several days. So, the used criteria (although widely used in the literature) seem to be quite subjective. It is not too difficult to imagine the geomagnetic activity as a continuous small storms release with very scarcely strong storms (similar to the way in which behave the Sun surface). From our results it is possible to envisage a probabilistic forecasting method. However, further scrutiny will be necessary. There are some works running on these lines and, given the difficulty mentioned at the end of the Data analysis and Preparation section, will be published, someday, elsewhere.

**Acknowledgements**

The authors sincerely acknowledge partial financial support from FAPERJ (Rio de Janeiro Founding Agency) and CNPq (Brazilian Founding Agency).

**Table Captions**

Table 1.- Storms that entered our classification. The column "Time" presents the time (UT) at which the storm started (when a value Dst $\leq -50$ nT was reached) and the column "Date" the dates of those events. The column "Depth" gives how much intense the storm was (first peak) and, finally, the column "Transient", the transient time before storm (the moment when the last value of Dst $\geq -10$ nT before the storm occurred).

**Figure Captions**

Figure 1.- Dst dependence for a period of one month. Beginning around the 27$^{th}$ day we can see some of the characteristics mentioned in the text: a rapid transient before storm and a slow recover after storm. Note that there are also some periods of Dst values > $-50$ nT (and that consequently do not entered our set, see the text) with the same characteristics (beginning around the 10$^{th}$ day, for example). The data was obtained from the World Data Centre for Geomagnetism at Kyoto.

Figure 2.- Distribution of $\Delta H$ values for the month of October 2000. There are two well-defined power law regimes.

Figure 3.- a) Distribution of $\Delta H$ values for three days (the three days anteceding the December 12, 2001 storm). The statistics is of worse quality than in Figure 2. Only one power law regime is more or less defined. b) Amplitude versus frequency of the Fourier transform for the same period.

Figure 4.- Time dependence of the slope of the distribution (in log-log plots). a) for ΔH; b) for the Fourier transform.

Figure 5.- Correlation between the slope during the three days immediately before storms and the slope for the three days before them. a) for ΔH; b) for the Fourier transform.

Figure 6.- Distribution of d values for periods near (circles) and far (squares) from storms. a) for ΔH; b) for the Fourier transform.

Figure 7.- Dependence of slopes for periods near (circles) and far (circles) from storms on the corresponding storm intensity. a) for ΔH; b) for the Fourier transform. Note the forbidden area in the lower left quadrant of a), less pronounced in b).

Table 1

| No. | Date (d/m/y) | Depth (NT) | Transient (hours) | Time (UT) | No. | Date (d/m/y) | Depth (nT) | Transient (hours) | Time (UT) |
|---|---|---|---|---|---|---|---|---|---|
| 1 | 18/02/98 | -100 | 10 | 01:00 | 32 | 13/10/00 | -71 | 5 | 06:00 |
| 2 | 10/03/98 | -116 | 7 | 21:00 | 33 | 10/11/00 | -96 | 5 | 13:00 |
| 3 | 21/03/98 | -85 | 6 | 16:00 | 34 | 27/11/00 | -80 | 3 | 02:00 |
| 4 | 24/04/98 | -69 | 7 | 08:00 | 35 | 23/12/00 | -62 | 6 | 05:00 |
| 5 | 02/05/98 | -85 | 8 | 18:00 | 36 | 13/02/01 | -50 | 4 | 22:00 |
| 6 | 14/06/98 | -55 | 6 | 11:00 | 37 | 19/03/01 | -105 | 9 | 22 |
| 7 | 26/06/98 | -101 | 4 | 05:00 | 38 | 28/03/01 | -56 | 4 | 02:00 |
| 8 | 06/08/98 | -138 | 8 | 12:00 | 39 | 11/04/01 | -271 | 8 | 24:00 |
| 9 | 18/09/98 | -51 | 4 | 14:00 | 40 | 18/04/01 | -114 | 5 | 07:00 |
| 10 | 25/09/98 | -170 | 4 | 05:00 | 41 | 18/06/01 | -61 | 4 | 09:00 |
| 11 | 19/10/98 | -109 | 10 | 13:00 | 42 | 17/08/01 | -105 | 4 | 22:00 |
| 12 | 25/12/98 | -57 | 4 | 12:00 | 43 | 13/09/01 | -57 | 5 | 08:00 |
| 13 | 24/01/99 | -52 | 7 | 23:00 | 44 | 23/09/01 | -55 | 5 | 16:00 |
| 14 | 18/02/99 | -123 | 6 | 10:00 | 45 | 19/10/01 | -57 | 10 | 22:00 |
| 15 | 01/03/99 | -94 | 6 | 01:30 | 46 | 24/11/01 | -221 | 10 | 17:00 |
| 16 | 29/03/99 | -56 | 3 | 15:00 | 47 | 21/12/01 | -67 | 8 | 23:00 |
| 17 | 17/04/99 | -90 | 5 | 04:00 | 48 | 30/12/01 | -58 | 3 | 06:00 |
| 18 | 31/07/99 | -53 | 4 | 02:00 | 49 | 10/01/02 | -51 | 10 | 22:00 |
| 19 | 22/09/99 | -173 | 3 | 24:00 | 50 | 02/02/02 | -86 | 7 | 10:00 |
| 20 | 22/10/99 | -237 | 7 | 07:00 | 51 | 01/03/02 | -71 | 5 | 02:00 |
| 21 | 11/01/00 | -81 | 6 | 22:30 | 52 | 17/04/02 | -98 | 6 | 18:00 |
| 22 | 23/01/00 | -97 | 7 | 01:00 | 53 | 11/05/02 | -110 | 6 | 20:00 |
| 23 | 06/04/00 | -287 | 6 | 23:00 | 54 | 19/05/02 | -58 | 4 | 07:00 |
| 24 | 16/04/00 | -60 | 9 | 06/00 | 55 | 01/08/02 | -51 | 3 | 14:00 |
| 25 | 24/04/00 | -61 | 4 | 15:00 | 56 | 04/09/02 | -109 | 3 | 06:00 |
| 26 | 17/05/00 | -92 | 5 | 06:00 | 57 | 24/10/02 | -69 | 5 | 06:00 |
| 27 | 24/05/00 | -147 | 8 | 09:00 | 58 | 20/11/02 | -87 | 4 | 21:00 |
| 28 | 08/06/00 | -90 | 4 | 20:00 | 59 | 27/11/02 | -64 | 8 | 07:00 |
| 29 | 26/06/00 | -53 | 8 | 12:00 | 60 | 19/12/02 | -71 | 7 | 19:00 |
| 30 | 15/07/00 | -289 | 6 | 22:00 | | Mean transient | | ~ 6 hours | |
| 31 | 12/09/00 | -73 | 10 | 20:00 | | Number of storms | | 60 | |

Figure 1

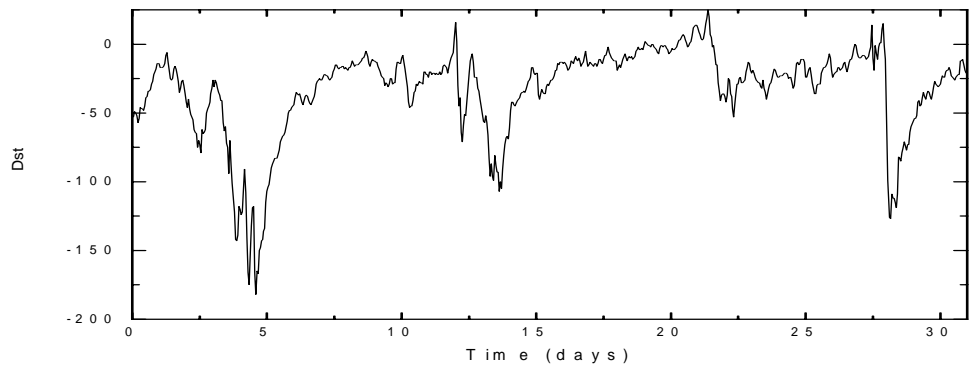

Figure 2

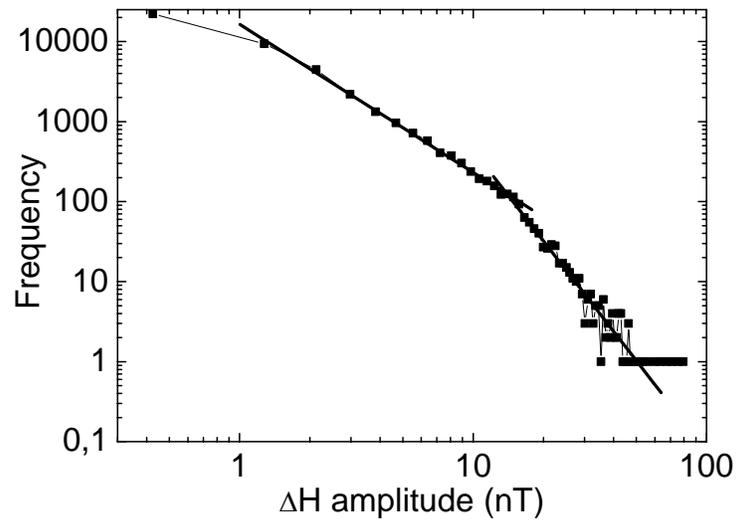

Figure 3

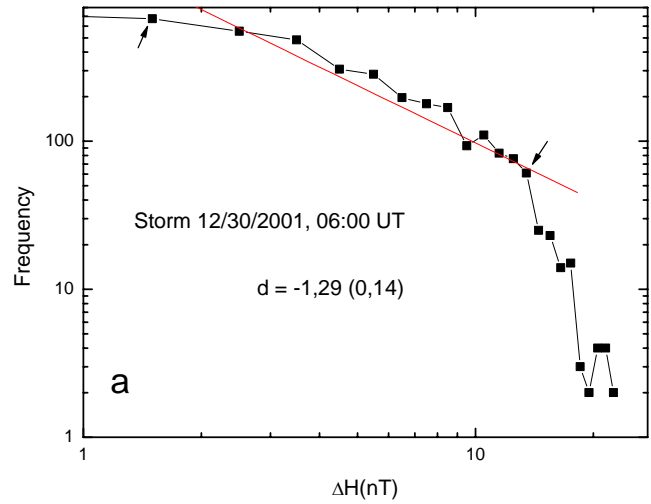

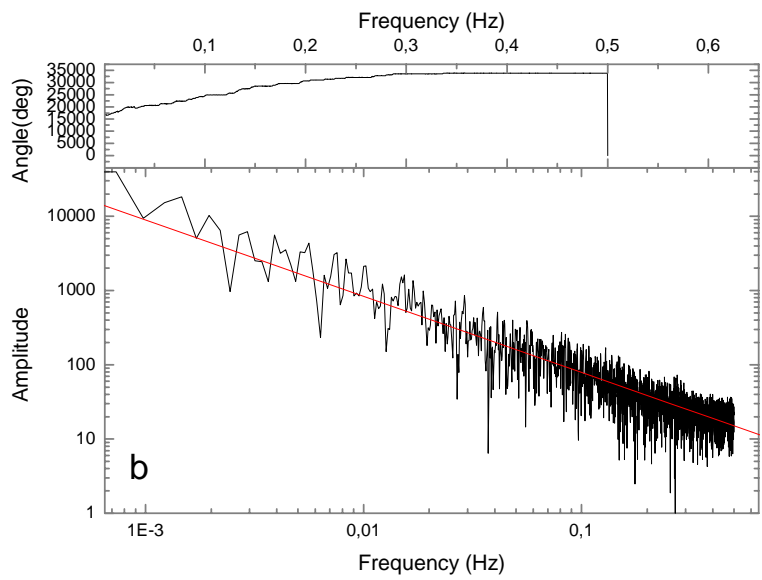

Figure 4

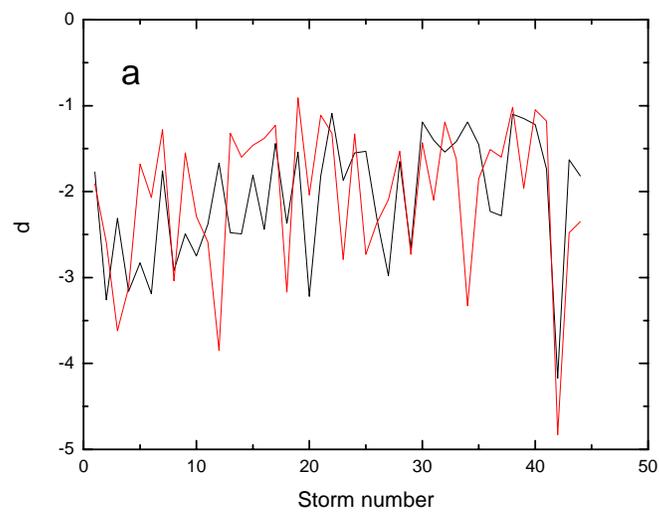

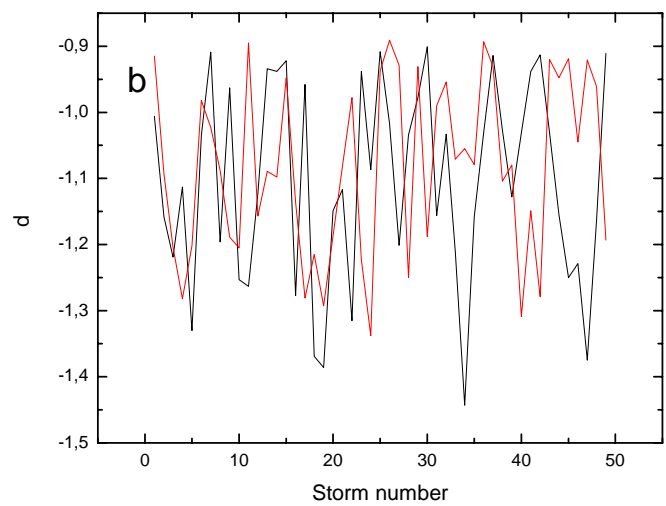

Figure 5

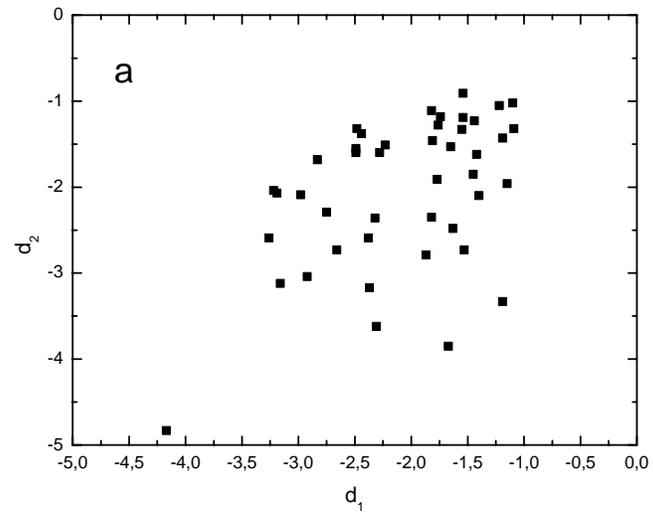

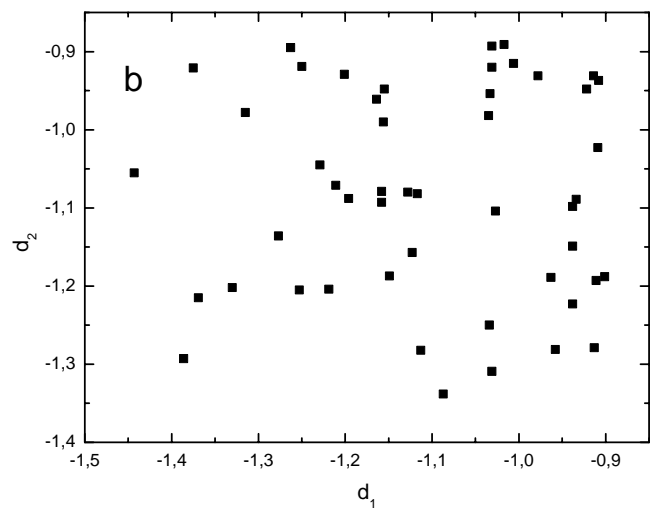

Figure 6

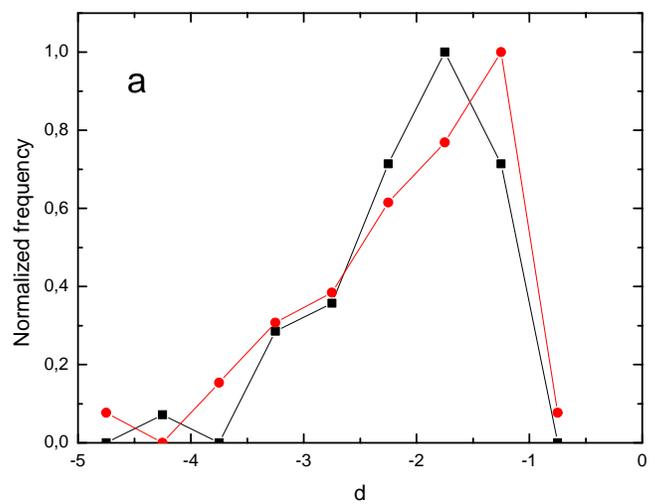

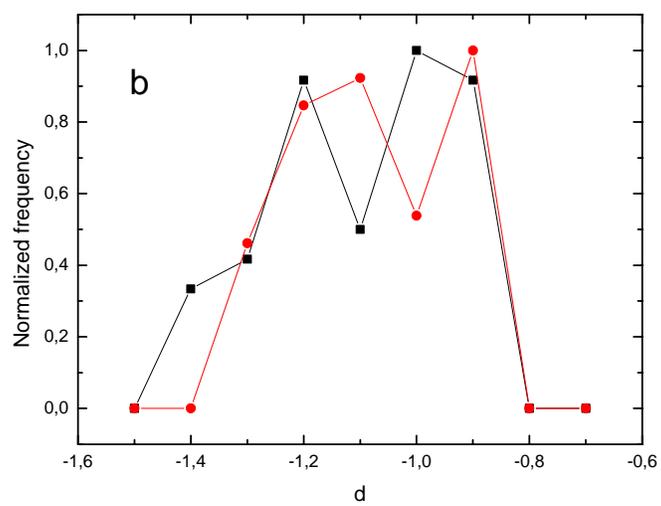

Figure 7

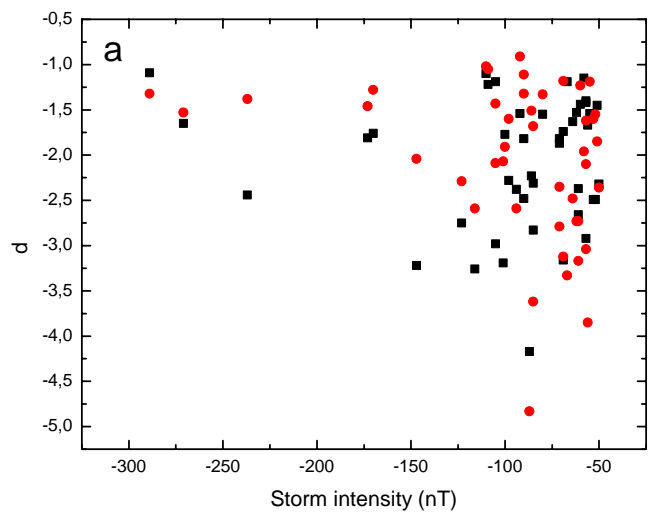

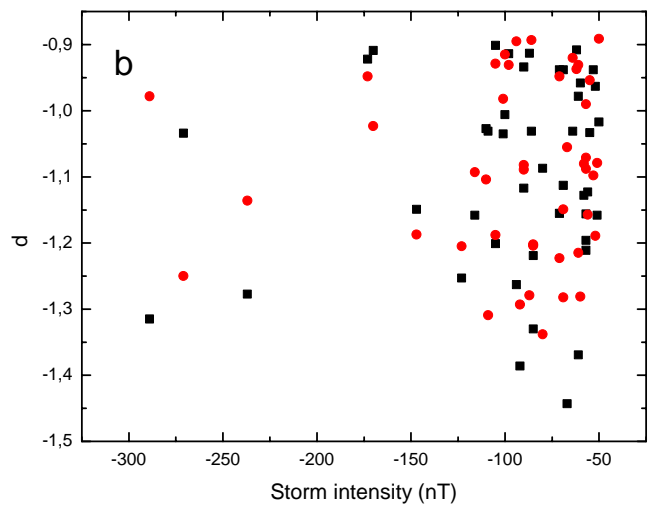